\UseRawInputEncoding
\documentclass[letterpaper]{article}
\usepackage{iccc}

\newcommand{\cameraready}{1} 

\usepackage{times}
\usepackage{helvet}
\usepackage{courier}
\usepackage{graphicx}
\title{Form Forge: Latent Space Exploration of Architectural Forms \\
via Explicit Latent Variable Manipulation}

\ifnum\cameraready=1
    \author{Kevin Dunnell, 
    Andy Lippman\\
    \ Media Lab\\
    \ Massachusetts Institutes of Technology\\
    \ Cambridge, MA 02142 \\
    \{dunnell, lip\}@media.mit.edu\\
    }
\else
    \author{Anonymous}
\fi
\begin{document} 
\maketitle

\begin{abstract}
This paper presents `Form Forge,' a prototype of a creative system for interactively exploring the latent space of architectural forms, inspired by François Blanciak's \textit{SITELESS: 1001 Building Forms} via direct manipulation of latent variables. Utilizing a fine-tuned StyleGAN2-ADA model, the system allows users to navigate an array of possible building forms derived from Blanciak's sketches. Distinct from common latent space exploration tools that often rely on projected navigation landmarks, Form Forge provides direct access to manipulate each latent variable, aiming to offer a more granular exploration of the model's capabilities. Form Forge's design is intended to simplify the interaction with a complex, high-dimensional space and to serve as a preliminary investigation into how such tools might support creative processes in architectural design.

\end{abstract}

\section{Introduction}

Latent spaces of generative AI models represent a compressed, abstract representation of the underlying training data. Exploring these spaces has emerged as a frontier for innovative design and creativity. While many recent advancements in text-to-image synthesis offer a means for expressing the desired outcome in language, they can abstract the underlying parameters of the model and inhibit the user from exploring the breadth of possibilities. Form Forge, a project inspired by François Blanciak's book \textit{SITELESS: 1001 Building Forms} (\citeyear{blanciak2008siteless}), presents a system for such exploration in the context of architectural forms. This work leverages StyleGAN2-ADA, a generative adversarial network (GAN), to extend the finite 1001 hand-sketched forms by Blanciak to a sea of similar synthetically generated architectural forms. Paired with dynamic front-end rendering, Form Forge allows users to interactively navigate through this latent space of potential architectural forms.

The interface of Form Forge features an innovative design where a circle of dynamic ticks surrounds the generated output image of a building form. Each tick is mapped to a specific latent variable of the StyleGAN model, enabling real-time manipulation and offering users a tangible way to see how adjustments in the latent space affect the generated forms. The project is a creative endeavor that seeks to reimagine how architects and designers engage with form and structure using machine learning-powered tools while speculating on novel approaches to exploring the latent spaces of generative AI models.

After the initial development of this project, we reached out to François Blanciak, who provided valuable feedback and guidance on the ethical use of his work, ensuring compliance with copyright considerations.

\begin{figure*}
    \centering
    \includegraphics[width=0.35\textwidth]{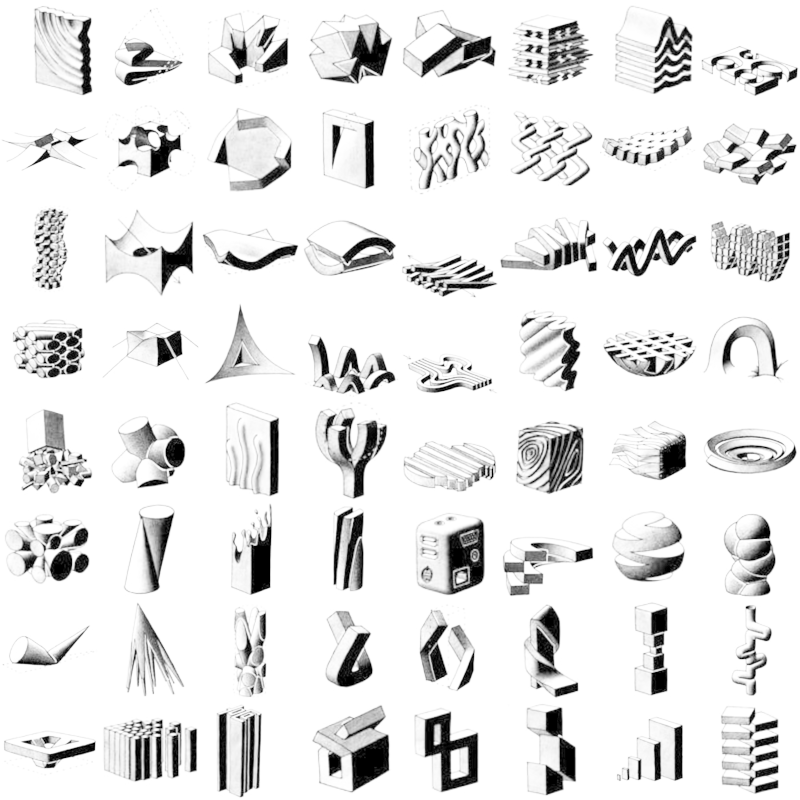}
    \caption{Organized square sample images from \textit{SITELESS: 1001 Building Forms} (\citeyear{blanciak2008siteless})}
\end{figure*}

\section{Related Work}

Form Forge intersects several critical areas of human-computer interaction (HCI), generative design, and machine learning. In crafting this work, we draw inspiration from and build upon a range of related research that informs and contextualizes our project.

\subsection{Latent Space Exploration}
Navigating latent spaces for creative purposes has been an area of growing interest. Previous works have explored the use of latent space manipulation to support the generation of novel images  \cite{Jahanian2020} \cite{DiPaola} \cite{Dang_2022}, user interfaces \cite{Mozaffari_2022}, mood boards \cite{10.1145/3563657.3596072}, and complex systems \cite{dunnell2023latent}. Others have extended these works to build systems that facilitate the disentanglement of latent variables \cite{Kahng_2019} \cite{Evirgen_2023}. These studies provide insights into how latent variables can be adjusted to yield diverse and innovative outcomes, an essential aspect of our project.

\subsection{Generative Adversarial Networks in Design}
GANs are increasingly being utilized across a wide range of creative and design industries for various innovative applications \cite{Hughes2021GANs}. Karras et al.’s foundational work on StyleGAN and StyleGAN2 provides the technical underpinning for our project \cite{Karras2019} \cite{Karras2020}. The release of StyleGAN2-ADA introduced a discriminator augmentation mechanism that made training GANs with smaller datasets feasible \cite{Karras2020StyleGAN2ADA}.

\subsection{HCI and Interactive Design Tools}
The HCI community has long been interested in developing tools that enhance user interaction with computational systems. Research by  Mark Gross and Ellen Do et al. on sketch-based interfaces in architectural design highlights the importance of intuitive and user-friendly tools for creative processes (\citeyear{Gross1996}). A survey on various Human-Generative-AI interaction approaches was completed by Jingyu Shi and Rahul Jain, et al (\citeyear{shi2024hcicentric}). Our project aims to explore how these findings can be applied to further enhance the user experiences of generative models in architectural design.

\subsection{Enhancing User Agency in Design}
Schneiderman emphasizes the importance of agency in digital tools in his `Eight Golden Rules of Interface Design’ by suggesting users should actively maintain control in shaping outcomes \cite{Schneiderman1998}. Our project aligns with this line of thinking by enabling users to directly manipulate underlying variables corresponding to design elements, thereby enhancing their sense of control and creativity.

By synthesizing these diverse strands, our project contributes a novel approach to design exploration, pushing the boundaries of user interaction and creative expression in architectural form design.

\section{Methodology}
Prototyping the Form Forge system entailed a multi-faceted process that involved the creation of a unique dataset, training a StyleGAN model, and developing an interactive interface. The system is designed to enable users to explore the latent space of architectural forms in a novel and interactive way.

\subsection{Dataset Creation}
Form Forge began with the creation of a dataset based on \textit{SITELESS: 1001 Building Forms} by François Blanciak (\citeyear{blanciak2008siteless}). Initially, there was an attempt to automate the process of systematically capturing images from a digitally purchased version of the book. However, the inconsistent layout and sizing of images in this version of the book presented a challenge for the rudimentary image scraper constructed. Alternatively, all 1017 sketches were manually screenshotted (16 additional sketches beyond the advertised 1001). This labor-intensive process ensured that each image was captured in a consistent 1:1 square format, initially at a resolution of approximately 128 pixels. To prepare the images for the StyleGAN training pipeline, the dataset was standardized by converting all images to a uniform style: white sketches on black backgrounds. This step was particularly crucial to address the issue of some sketches being rendered with inverted colors in the digital copy. Finally, the images were upscaled to 512x512 pixels for consistency with the target output resolution. 

\subsection{Dataset Visualization}
As a sidestep in the process of constructing our system and to better understand the variety of existing forms available for training the StyleGAN model, the dataset was processed using the Inception V3 convolutional neural network (CNN) to extract a 1-dimensional vector representation of each image \cite{szegedy2016rethinking}. Subsequently, t-SNE (t-distributed Stochastic Neighbor Embedding) was employed to reduce the dimensionality to two dimensions, facilitating the clustering of visually similar entities \cite{van2008visualizing}. To optimize the spatial arrangement of the data points in the t-SNE plot, the Laplacian-Jordan-Von Neumann (LapJV) algorithm was applied \cite{jonker1988shortest}. This algorithm minimizes the total displacement of each point by optimally assigning them to positions on a predefined grid. The resulting organized square samplings of the dataset are illustrated in Figure 1 during the initial stages of the data collection process.

\subsection{StyleGAN Model Fine-tuning}
Form Forge employs an instance of StyleGAN2-ADA that has been fine-tuned on the training dataset derived from \textit{SITELESS: 1001 Building Forms} (\citeyear{blanciak2008siteless}). This collection of images was augmented using horizontal mirroring, effectively doubling the dataset size from 1017 to 2034 images while avoiding y-mirroring due to the directional nature of some sketches. The model underwent 235 training ticks, corresponding to approximately 461 epochs, given the processing of 4,000 images per tick. This training, conducted with standard parameters including a learning rate of 0.002 as recommended by Nvidia's StyleGAN2-ADA-PyTorch Google Colab notebook \cite{Karras2020StyleGAN2ADA}, allowed the model to iteratively refine its output, accurately capturing the essence of Blanciak’s architectural designs in both static images and animations.

  \begin{figure*}
    \centering
    \includegraphics[width=0.81\textwidth]{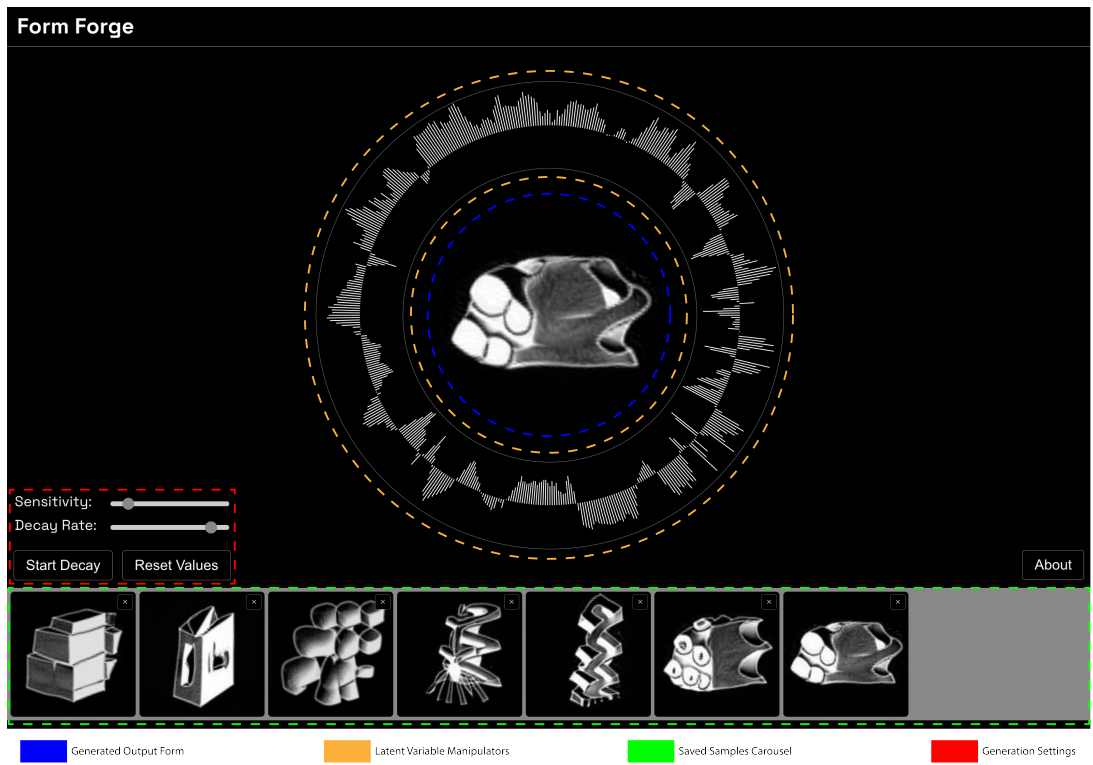}
    \caption{Form Forge User Interface}
  \end{figure*}

\subsection{Interactive Interface Development}
The culminating component of our system was the creation of a web-based interactive interface. The system was developed using Flask to support StyleGAN model inference on the back end and utilizing React for responsive and user-friendly front-end design. 

\subsubsection{Back-end Generation Server}
A back-end Python server hosts the model and supports the near real-time sampling of latent space, given an array of user-adjusted latent variables. A portion of the Nvidia Google Colab notebook, specifically the section `Testing/Inference $>$ Generate Single Images,' was adapted to create an image by taking an array of 512 latent variables\cite{Karras2020StyleGAN2ADA}. This functionality was then incorporated into the back-end Flask server function, which is the backbone of our web-based interactive interface.

\subsubsection{Generated Output Form}
At the core of the interface is the generated architectural form, prominently displayed in the center (delineated by blue dashed lines in Figure 2). This image, sized at 512x512 pixels, represents a form decoded from the model's latent space, specifically at the 512-dimensional vector location determined by the user. The visualization serves as a direct reflection of user adjustments within the system's interactive environment.

\subsubsection{Latent Variable Manipulators}
Encircling the central output image are 512 dynamic tick marks, each corresponding to one of the 512 dimensions of the StyleGAN model's latent space (between the pair of orange dashed lines in Figure 2). User interactions with this component involve cursor movements: moving the cursor outside the ring of ticks increases its length in alignment with the cursor's angle, whereas moving the cursor inside the ring reduces the length inversely. This mechanism is designed to visually represent the relative adjustments made to each latent variable. Two thin boundary lines, both outside and inside the tick marks, define the limits of this interaction, preventing constant flux in tick adjustments as users navigate other interface elements.

\subsubsection{Saved Samples Carousel}
As users navigate through the latent space, they can save forms that appeal to them by left-clicking anywhere on the screen. This action captures both the visual output and the underlying vector representation, storing them in the "Saved Samples Carousel" located at the bottom of the interface (identified by the green dashed lines in Figure 2). Users can restore saved images to the central display area and their corresponding tick configurations by clicking on an image in the carousel. Additionally, users can remove images from the carousel by selecting the `X' at the top right corner of each thumbnail.

\subsection{Generation Settings}
Positioned in the bottom-left corner of the screen are several adjustable settings that enhance user interaction (red dashed lines in Figure 2). These include a decay function toggle, which gradually resets all tick marks and their values to zero, providing an animated transition back to the latent space's origin. An immediate reset button also exists to instantaneously revert all latent variables to their initial zero values. Accompanying sliders allow users to control the decay rate when the decay function is active and adjust the sensitivity of how mouse movements influence the rate of change in tick lengths and latent variable values.

\section{Discussion}
Form Forge introduces a focused approach to exploring the latent space of architectural forms derived from François Blanciak's \textit{SITELESS: 1001 Building Forms} (\citeyear{blanciak2008siteless}). By enabling direct manipulation of latent variables in a fine-tuned StyleGAN2-ADA model, this system provides a novel way for users to interact with architectural possibilities, potentially offering new insights into design methodologies and HCI. While this was a creative, exploratory project, we are excited about how this work may contribute to designers and architects interacting dynamically with latent spaces of generative models.

To systematically evaluate user interaction with Form Forge, we suggest developing a set of metrics that could include task completion times, error rates, user satisfaction surveys, and qualitative feedback on creativity enhancement. An initial user study could employ these metrics to assess the interface's intuitiveness and the interaction model's effectiveness in fostering an engaging and creative user experience.

\subsection{User Experience and Interaction}
Central to our project is the ability to explicitly manipulate individual latent variables to generate images of architectural forms from latent space. The interface provides feedback on the extent of the adjustments made to a set of latent variables and how these perturbations influence the resulting output image. However, preliminary tests with the system exposed challenges with this method of interaction.

A fundamental limitation is that the connection between manipulating individual ticks and their impact on the displayed form is not inherently intuitive. This challenge stems from the high-dimensional nature of the latent space, where relationships between variables and visual outcomes are complex and not yet disentangled for this specific dataset. As a result, users might find it difficult to predict how their interactions will alter the generated forms.

Addressing this issue requires implementing techniques for latent disentanglement, which could make the relationships within the latent space more understandable and controllable. This enhancement could significantly improve user agency and control, making the design process more intuitive and accessible.

\subsection{Potential Developments and Future Work}
The development of Form Forge opens several avenues for future enhancements and research. A primary area for improvement is the enhancement of image quality. The resolution of images, currently upscaled from 128x128 to 512x512 pixels, directly affects the visual clarity and detail of the architectural forms, which can impact user satisfaction and the system's overall usability. To enhance this aspect, we propose integrating super-resolution upscaling models before training the StyleGAN. This approach is expected to significantly improve the fidelity and detail of generated outputs, thereby enhancing the user's ability to engage with and evaluate complex forms.

Another exciting avenue for extending our tool is to support the generation of 3D forms, moving beyond the current rendering of 2D sketches. Enabling users to explore and manipulate three-dimensional architectural designs directly would significantly enhance our system's creative possibilities and applicability. 

In light of recent advancements, exploring next-generation generative models like StyleGAN3 or diffusion models could offer substantial benefits. StyleGAN3, for instance, provides improved alignment and consistency in generated forms, which could enhance the visual continuity of architectural designs. Similarly, diffusion models are known for their superior sample diversity and quality, attributes that could further expand the creative potential of Form Forge. A comparative analysis of these models' capabilities will guide their potential integration into our system.

Lastly, the project would benefit from a structured evaluation metric to assess user interaction, engagement, and the effectiveness of the interface in facilitating creative exploration. Such an evaluation could provide valuable insights for refining the system and understanding its impact on the design process.

\section{Conclusion}

Form Forge explores the integration of HCI, AI, and architectural design, demonstrating the potential for manipulating latent spaces in creative processes. This project introduces interactive techniques for navigating and manipulating high-dimensional latent spaces to steer the outputs from generative models. Nevertheless, it also highlights several challenges and areas for further development. Looking forward, we seek to engage with the HCI community to receive feedback on our proposed evaluation methods. Such collaboration will be crucial in refining our system and improving its interface for exploring latent spaces in generative models. We aim to enhance the capabilities of users, architects, and designers by providing them with greater control and agency in their design explorations. This paper has attempted to lay a foundation for such enhancements and invites the community's input to guide our ongoing improvements.

\section{Acknowledgements}

I want to thank Zach Lieberman for his guidance and inspiration through his course \textit{Drawing++}, which led to the initial concept for this prototype. Thanks to François Blanciak, the author of \textit{SITELESS: 1001 Building Forms}, his compelling collection of images sparked the imagination to envision the possibilities between his drawings. François' sketches provided the essential foundation for this work and he helped refine the scope of image usage to comply with copyright requirements. I also sincerely appreciate Caitlin Mueller's encouragement in publishing this project. Lastly, I acknowledge ChatGPT for assisting in the creation of this publication. 

\bibliographystyle{iccc}
\bibliography{iccc}

\end{document}